# Simulated Stress and Stretch of SWCNT

Ana Proykova[*] and Hristo Iliev
University of Sofia, Faculty of Physics
5 James Bourchier Blvd, Sofia 1126, Bulgaria

## Abstract

The mechanical stability of open single wall carbon nanotubes (SWCNT) under axial stress (compression and tension) and twist has been re-examined in a search of specific tube-length and load scaling. SWCNT with different chiralities and lengths have been simulated with a classical molecular dynamics method employing the many-body empirical Tersoff-Brenner potential. Stress has been achieved by enforcing constant linear velocity on the edge atoms from both sides of the SWCNT as suggested by Srivastava and Barnard. We have found opposite length scaling at fast (1/10 of $v_S$ the sound velocity in carbon tubes) and slow (1/20 $v_S$) loading of (10, 0) tubes. Another finding is that at fast loading short zigzag (10, 0) tubes transform from elastic to plastic states before they break in the middle, while tubes, longer than 13 nm, break-up directly in the elastic state. Thus, short tubes behave like metals or ionic solids, while long tubes resemble ceramics or glasses under the conditions studied. All tubes form spiral-like structures when twisted. Standing waves, generated under specific conditions, determine the different behavior of tubes with various lengths and chiralities.

## Introduction

Since their discovery [1], single-wall carbon nanotubes (SWCNT) have been considered as members of the fullerene family [2], having a high aspect ratio, a few defects, and unique mechanical and electronic properties. Their mechanical properties, reviewed in [3], are closely related to both electron conductivity [4] and adsorption [5]. Hence, mechanical loads on SWCNTs can be used to design a good dynamical chiller, based on adsorption–de-sorption of water molecules. This possibility of engineering justifies new simulations of loading, including twist of tubes, which is motivated by a possible application of nanotubes as shafts in nano-electromechanical devices.

Nanotubes are quasi-3D cylindrical objects made of rolled-up graphite sheets, Fig.1. The vector $\mathbf{C}_h = n\mathbf{a}_1 + m\mathbf{a}_2 \equiv (n, m)$; $n, m$ - integers, connects crystallographically equivalent sites on the sheet. The angle $\theta$ between $\mathbf{C}_h$ and the zigzag line (n,0) specify the tube type: (2n,n) is chiral, (n,n) is armchair, (n,0) is zigzag. The tube diameter is computed as follows: $d=0.078 (n^2 + m^2 + nm)^{1/2}$ nm. The tube curvature and chirality determine the SWCNT conductivity. The zigzag (n,0) SWNTs should have two distinct types of behaviour: the tubes will be metals when $n/3$ is an integer, and otherwise semiconductors. However, the tube (5,0) with $d < 4$ Å is metallic, which should be related to the band-gap change due to the curvature [6].

---

[*] Corresponding author. Phone: +359 2 8161828, Fax:+359 2 962 52 76, E-mail: anap@phys.uni-sofia.bg

This article presents evidences for different responses of tubes with different lengths to applied stress and twist: short ($l < 13$ *nm*) zigzag (10,0) tubes sustain a large amount of compression and recover their initial shape when the external force is set to zero. The longer tubes fold irreversibly under the same force (strain energy per atom).

We simulate the behavior of open-ended SWCNTs (no periodic boundary conditions) under heavy loading. We consider zigzag (10,0), armchair (6,6), and (10,10) tubes with lengths between 8,63 *nm* and 54,3 *nm*. The aim is to answer the following questions: What is the amount of stress (twist) necessary to *break bonds*, i.e. elastic-to-plastic transformation of a SWCNT with a given length and chirality? What is the change of the potential energy surface (PES) under such extreme loading?

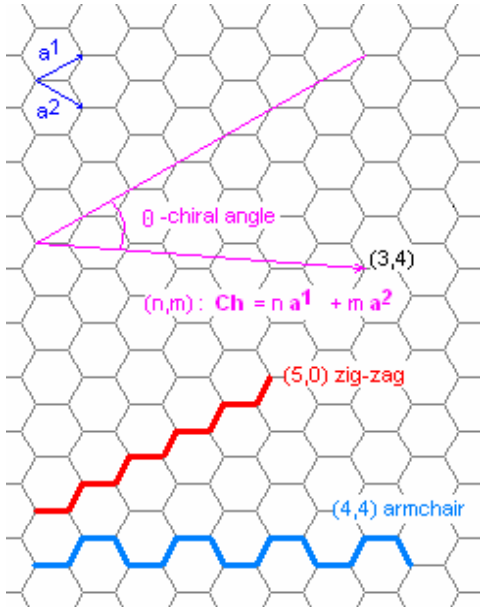

Fig.1 Rolled-up graphite sheets make tubes.

The first question arises in relation of elastic-to-plastic transition to the internal atomic response to an externally applied stress. The carbon atoms in tubes are covalently bonded - the stiffest bond that determines the exclusive mechanical stiffness of carbon nanotubes. However, a given material has two kinds of stiffness: one for a fast loading, when there is too little time for relaxation, 'unrelaxed' process, and another for slow loading which allows relaxation to occur, relaxed process. In solids with dense packing these two do not differ much. Here we show how carbon nanotubes behave under fast and slow loading.

The second question is related to physisorption: a PES with many local minima is a better adsorbent than an even one. Broken bonds, due to high stress, facilitate creation of adsorbing areas in PES.

Up to our knowledge, the scaling of tension induced plasticity of carbon nanotubes is studied here for the first time.

## Model for mechanical loading

To model formation of carbon tubes, we solve the Newtonian equation of motion $m_i \ddot{\mathbf{r}}_i(t) = -\nabla_{\mathbf{r}_i} E_b$, with $E_b$ being a simplified form of the Brenner-Tersoff potential [7]:

$$E_b = \sum_i \sum_{j(>i)} \left[ V_R(r_{ij}) - \overline{B}_{ij} V_A(r_{ij}) \right], \quad (1)$$

$i$ and $j$ run over all atomic sites; $r_{ij}$ is the distance between *i*-th and *j*-atom; the repulsive term $V_R$ is

$$V_R(r_{ij}) = f_{ij}(r_{ij}) D_{ij}^{(e)} / (S_{ij} - 1) e^{-\sqrt{2 S_{ij}} \beta_{ij} (r_{ij} - R_{ij}^{(e)})} \quad (2)$$

and the attractive terms $V_A$ is

$$V_A(r_{ij}) = f_{ij}(r_{ij}) D_{ij}^{(e)} S_{ij} / (S_{ij} - 1) e^{-\sqrt{2/S_{ij}} \beta_{ij} (r_{ij} - R_{ij}^{(e)})}, (3)$$

with $D_{ij}^{(e)}$, $S_{ij}$, $\beta_{ij}$ and $R_{ij}^{(e)}$ - parameters given in Table 1. The function $f_{ij}$, which restricts the pair potentials $V_R$ and $V_A$ to nearest neighbors is:

$$f_{ij}(r) = \begin{cases} 1, r < R_{ij}^{(1)} \\ \left[ 1 + \cos\left[ \frac{\pi(r - R_{ij}^{(1)})}{R_{ij}^{(2)} - R_{ij}^{(1)}} \right] \right] / 2, R_{ij}^{(1)} < r < R_{ij}^{(2)} \\ 0, R_{ij}^{(2)} < r \end{cases}$$

$\overline{B}_{ij} = (B_{ij} + B_{ji})/2$ is an empirical bond-order function; each of the averaged terms has the form:

$$B_{ij} = \left[1 + \sum_{k(\neq i,j)} G_i(\phi_{ijk}) f_{ik}(r_{ik}) e^{\alpha_{ijk}\left[(r_{ij}-R_{ij}^{(e)})-(r_{ik}-R_{ik}^{(e)})\right]}\right]^{-\delta_i}$$

For carbon atoms (index 'C'), the function $G_C$ is:
$$G_C(\phi) = a_0\left\{1 + c_0^2/d_0^2 - c_0^2/\left[d_0^2 + (1+\cos\phi)^2\right]\right\}$$

$\phi$ is the angle between the lines, connecting $i$-th with $j$-th atom and $i$-th with $k$-th atom. The potential parameters have been adjusted [7] to reproduce the bonding structure of graphite, diamond, carbon nanotubes and small hydrocarbon molecules.

| Parameter | Value | Parameter | Value |
|---|---|---|---|
| $R_{ij}^{(e)}$ | 1.39 Å | $R_{ij}^{(1)}$ | 1.7 Å |
| $D_{ij}^{(e)}$ | 6.0 eV | $R_{ij}^{(2)}$ | 2.0 Å |
| $\beta_{ij}$ | 2.1 Å$^{-1}$ | $a_0$ | 0,00020813 |
| $S_{ij}$ | 1.22 | $c_0^2$ | $330^2$ |
| $\delta_{ij}$ | 0.5 | $d_0^2$ | $3.5^2$ |
| $\alpha$ | 0.0 | | |

**Table 1**: Potential parameters used in simulations.

We use the Molecular Dynamics (MD) method [8], which is a numerical technique for integration of ordinary differential equations. In the presence of an external field $F$ (stress or twist) applied to a collection of atoms, the Newton's second law is:
$$m_i \ddot{\mathbf{r}}_i(t) = -\nabla_{\mathbf{r}_i} E_b + \mathbf{F}_{ext}(t), \qquad (1)$$

$m_i$ is the $i$-th atom mass. Integration is performed with the velocity Verlet algorithm [8], which advances in time the particle positions and velocities as follows:

$$\begin{vmatrix} \mathbf{r}_i(t+dt) = \mathbf{r}_i(t) + dt\cdot\mathbf{v}_i(t) + (2m)^{-1}dt^2\mathbf{F}_i(t) \\ \mathbf{v}_i(t+dt) = \mathbf{v}_i(t) + (2m)^{-1}[\mathbf{F}_i(t) + \mathbf{F}_i(t+dt)] \end{vmatrix},$$

where $dt$ is the time step (0.2 fs in our simulations). The $dt$ value satisfies two requirements: a) energy conservation $\Rightarrow dt$ must be small; b) reversibility of the classical trajectory if $dt \to -dt$, $\Rightarrow dt$ must be large enough to reduce the round-off errors accumulated in very long runs needed for a very small step. The value of $dt$ =0.2 fs ensures the energy conservation (in a constant energy run, no external field) up to $10^{-5}$ eV/atom. The cell linked-list method and the Verlet neighboring list [8] have been combined to speed-up the calculations [9].

The tube starting configurations – zigzag, chiral, or armchair - are generated with the Mintmire's code [10] and consequently relaxed (optimized) by a power quench in a MD run [11]. In the power quench, each velocity component is set to zero if it is opposite to the corresponding force of that component. This affects atomic velocities, or unit-cell velocities (for cell shape optimizations).

The stress is simulated by changing at each time step only the positions {$\mathbf{r}$} of the edge atoms of the open-end tube, {$\mathbf{r}$}→{$\mathbf{r} \pm \delta\mathbf{r}$} [12]. Hence, the edge atoms do not move according to the classical equation of motion, while positions and momenta of the rest of the atoms are computed from Eq. 1. The coordinate system is centered in the middle of the tube with the $z$ axis along the tube length. Thus, for compression the sign is (+) for the left side atoms and (-) – for the right side atoms. For tension, the signs are altered. Various values of constant velocity of the edge atoms have been tested to check the nanotube responses to fast and slow loading. Here we report results for 500 and 1000 m/s, e.g. 1/20 and 1/10 of the sound of velocity in carbon. The stress waves, corresponding to these velocities, cause (or not) a generation of standing waves depending on the tube length. Due to the standing waves and beating, some atoms of the tube do not displace while vibrate with maximum amplitudes, which produce bond-breaks.

To be exact, the stress σ at a point is determined by $\sigma = \lim_{\Delta A \to 0} \frac{\Delta P}{\Delta A}$, where ΔP is the load being carried by a particular cross section ΔA. In general, the stress may vary from point to point as it is in the case of non-equilibrated tubes we study. The stress has two components, one in the plane of the area A, the *shear stress*, and one perpendicular, the *normal stress*. It is always possible to transform the co-ordinates on the body into a set in which the shear stress vanishes. We consider the remaining normal stresses called the principal stresses. The stress cannot be measured directly but is usually inferred from measurements of strain, i.e. a change in size and/or shape. *Stretch* is the change in length: $e = \frac{\Delta l}{l_0}$, where $\Delta l$ is the change in length and $l_0$ is the original undeformed length. In tension $e>0$, i.e. the body has been lengthened; $e<0$ in compression.

The twist is realized as follows: at every time step we rotate in the opposite directions the edge atoms around the tube axes at angles between 0.001 and 0.00001 rad/s. Depending on the velocity of rotation standing waves again appear.

To study the effect of the standing and reflected waves we applied the external force at times 0, τ, 2τ,…:

$$m_i \ddot{\mathbf{r}}_i(t) = -\nabla_{\mathbf{r}_i} E_b + \sum_{n=0} \delta_{t,n\tau} \mathbf{F}_{ext}(t), \qquad (2)$$

where $\delta_{t,n\tau}$ is the Kronecker delta. For large values of τ (>2 ps), the tubes have time to relax between the successive loads and obey the Hook's law.

## Results

Defect formation (bond breaking, vacancy), energy accumulation, and shear of the tubes have been monitored during the calculations.

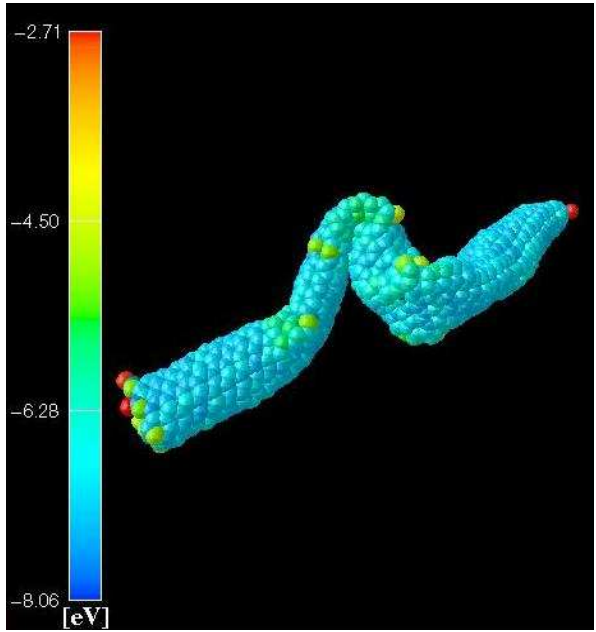

**Figure 2:** A short tube, (10,0) 12.98 *nm*, bends under compression.

Our results for uniaxial stress are consistent with the data published [12] for (10, 10) tubes with different lengths. Short tubes (up to 13 *nm*) sustain large amount of compression (~ 30%) without breaking and almost completely recovered their initial shape when the external force is set to zero.

Under compression several atoms reside off-plane Fig.2 and can only break one or two bonds. Less bounded atoms (lower binding energy) are green in the Figure 2. The red atoms are 'hot' and occasionally sublimate. If the compression time is short, the pressure is below 100 GPa, the tube completely restores its initial shape when released; otherwise it buckles. Salvetat and co-workers [3] have already shown in their experiments that the tubes are elastic for low loads. The elastic regime is proper for periodic adsorption/de-sorption of noble gases, which do not form a chemical bond with the dangling bonds. However, the rate of adsorption is too low because the number of defects created in compression is small – green-yellow spots in the figure.

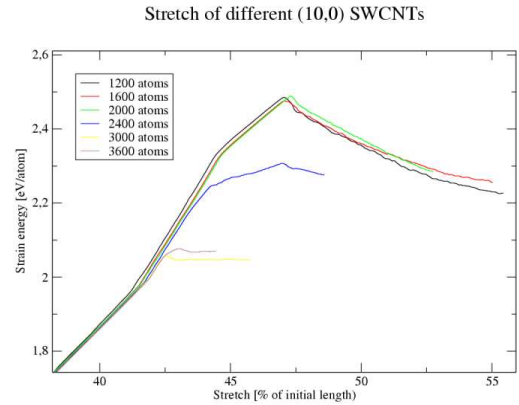

**Figure 3:** In tension two kinks of the curves are seen for short tubes; one kink - for long tubes. The maximum of the curves in all cases correspond to the break of tubes. The strain energy per atom rapidly reduces after the break.

Tension of the tubes, Figure 3, generates more defects then compression. Under tension, the tubes break for 42-44% extension of the different tubes studied here. This high percentage of extension is due to the covalent bonding of the carbon atoms and hexagonal structure of the graphene sheets. Inspecting the curve slopes, we notice two kinks for (10,0) tubes with lengths less than 26.03 *nm* (2400 atoms). We identify the first kink as transition between the elastic and plastic states with broken bonds that do not re-connect when the external force is set to zero. These dangling bonds fluctuate slowly and attract noble gases.

We relate the second kink with the appearance of a 'bridging' area in the middle of tubes shorter than 26,03 *nm* like in the Figure 4. The tube is in a plastic state. We call a state 'plastic' if the system does not restore its initial shape after the force is set to zero.

Bonds break when enough energy is accumulated under tension. The time $t$ (in *ps*), needed for a tube to gain energy for elastic-to-plastic transition, depends linearly on the tube length. The slow tension is given with $t(N) = 0,0102 N – 1,6626$ and the fast tension is $t(N) = 0.0045 N + 0.2612$; $N$ is the number of carbon atoms. The different slopes point to different redistribution of the energy in slow and fast processes.

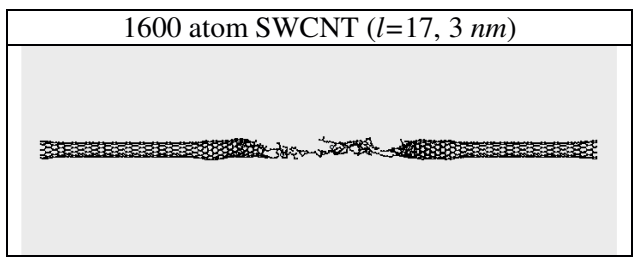

1600 atom SWCNT (*l*=17, 3 *nm*)

**Figure 4:** Broken bonds in the middle of the tube – a plastic state, corresponding to the second kink of the curve in the Figure 2.

Long tubes form 'bridges' nearby the two ends, Figure 5, and break directly from the elastic state.

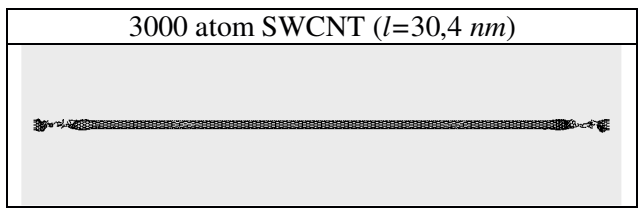

3000 atom SWCNT (*l*=30,4 *nm*)

**Figure 5**: Long tubes break in the elastic state: no elastic-to-plastic transition occur.

These results indicate a size effect: under tension, *short tubes behave like metals*, while long tubes resemble ceramics under mechanical loads. The reason is the energy distribution along the tube axis and the stress wave propagation, which we study in detail for fast and slow processes.

We report here different responses of the tubes to slow and fast tension: the longer the tube the larger extension before break-up under low speeds, while at the faster velocity, the length scaling is opposite.

| zigzag | No. of atoms | Tension velocity | Tension time | % stretch |
|---|---|---|---|---|
| (10,0) | 1200 | 500m/s | 10,36ps | 43,8 |
| (10,0) | 1200 | 1000m/s | 5,60 ps | 43,1 |
| (10,0) | 1600 | 500 m/s | 15,28ps | 44,1 |
| (10,0) | 1600 | 1000m/s | 7,52 ps | 43,4 |
| (10,0) | 2000 | 500 m/s | 18,20ps | 44,2 |
| (10,0) | 2000 | 1000m/s | 9,40 ps | 43,3 |
| (10,0) | 3000 | 500 m/s | 29,00ps | 44,5 |
| (10,0) | 3000 | 1000m/s | 13,76ps | 42,3 |

**Table 2**: Fast and slow tension scale oppositely.

Fast loading, i.e. unrelaxed stress process, causes elastic-to-plastic transition in short tubes before they break. Slow loading does not initiate elastic-to-plastic transition for any tube length studied here. We explain this with the tube- and the stress-wave lengths becoming incommensurate.

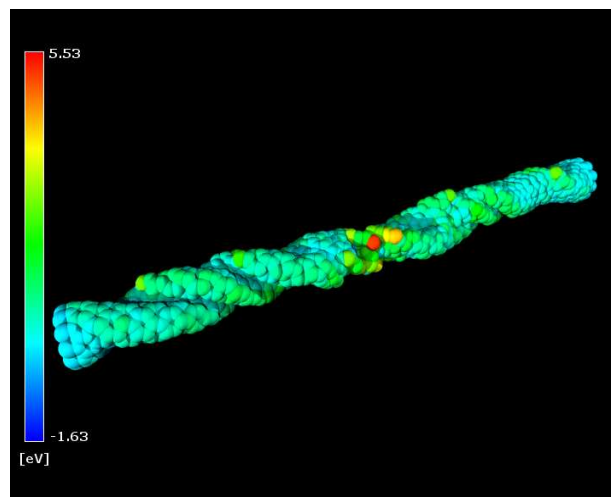

**Figure 6**: The red and yellow atoms are less bound in a twisted (10,0) tube with *l*=12,9 *nm*.

The scale in the Figure 6 corresponds to the difference between the initial energy and the current energy of every atom. The snapshot corresponds to 3,4 *ps* duration of twist. When off-line atoms occur, the twisted tube can not restore its original shape even if the external force is set to zero. We have periodically applied and released the twisting external force on the tube to study the process of energy accumulation and standing wave formation as a function of the period τ, Figure 7.

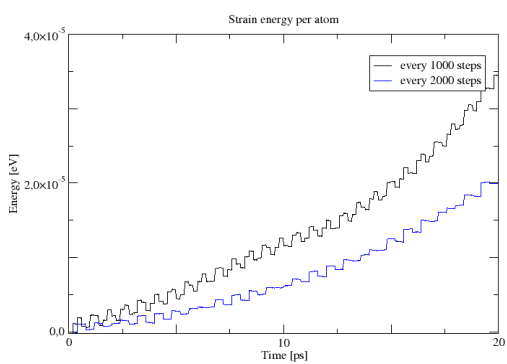

**Figure 7**: Accumulation of strain energy in a tube as a function of the period τ in Eq.2. Beating is well seen for τ =2000 steps = 0.4 *ps*: at 6, 10, 14, 18 ps.

The standing wave (beating) is a result of modulation of the stress-wave by reflected waves. In order to monitor the motion of each atom due to the standing wave, we perform Fourier transform of the atom kinetic energy. The Figure 8 shows a typical pattern of beating (~1.8 $10^{13}$ Hz) observed for a single atom located near the tube end. The pictures for atoms, located at different sites of the tube, are similar, although the peaks positions could be shifted due to the resonance conditions.

## IV. Conclusions and comments

This study shows that a relatively small number of dynamical defects can be created in an initially perfect SWCNT before it breaks-up. This is discouraging for the producers of defected nanotubes considered more interesting for nano-electronic applications. The good news is that once created, these defects do not migrate: a detailed study of defect formation and diffusion will be published in [15]. The atoms, surrounding a defect, vibrate and might become dynamical adsorption centers, which will be studied with the code based on the density-functional theory [13]. Our current calculations are limited as they are based on classical approximations.

In practice, the SWCNTs grow in ropes or bundles hence such computations could be of interest as well. However, these computations are expensive and require intelligent and new approaches. Possible solutions are the usage of clusters of computers and/or parallel computations as we have already demonstrated [9]. Our algorithm scales with the number of atoms N in a system as *O*(N).

We show that fast tension causes metal-like mechanical behavior of short tubes – they firstly transform from elastic to plastic state before break. Long tubes resemble ceramics or glasses under fast stress, i.e. break directly in the elastic state. This result should be remembered when designing nano tube devices.

Under tension, the atoms increase their temperature (zero in the beginning of the simulations). One could think of temperature induced transformations and related the specific size effects in mechanical load with the temperature-driven solid-solid changes in clusters of rigid molecules. These clusters exhibit various structural phase changes depending on the topography of the potential energy surface (PES) and particularly on the distribution of the saddle points on the PES [14]. Special research of the potential energy topography of elongated and twisted tubes will be performed in the future to study unrelaxed processes.

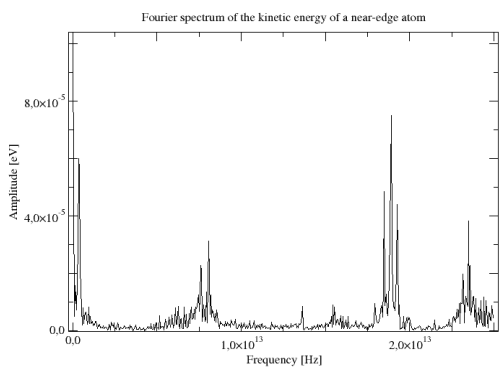

**Figure 8**: Beating is clearly seen in the Fourier spectrum of the kinetic energy of a single atom, $3^{th}$ ring from the edge atoms.

**Acknowledgements**
This research was supported by the Ministry of Education and Science (Bulgaria) under the grant (F-3/2003). H. Iliev acknowledges the support from the Edinburgh Parallel Computing Center during his visit in 2003 {EC contract No.HPRI-CT-1999-00026 TRACS-EPCC}. A.P. is very grateful to Prof. Jeffery Gordon (Ben-Gurion University of the Negev, Israel) and Dr. Chua (the University of Singapore) for their valuable ideas, inspiring discussions, and stimulating experimental work in the field of adsorption of carbon tubes.